\documentclass{mem}
\usepackage{natbib}\usepackage{txfonts}\usepackage{balance}
\usepackage{graphicx}
\usepackage[a4paper,breaklinks,dvipdfm]{hyperref}
\idline{75}{1}
\begin{document}
\def\teff{$T\rm_{eff }$}
\def\kms{$\mathrm {km s}^{-1}$}

\title{Evolutionary stellar models for multiple population Globular Clusters:}

\subtitle{theoretical framework and interpretative analysis}

\author{S. \,Cassisi\inst{1}, M. \,Salaris\inst{2}, and A. \,Pietrinferni\inst{1}}

  \offprints{S. Cassisi}

\institute{
$^1$INAF -- Osservatorio Astronomico di Teramo, Via M. Maggini, sn.,
I-64100 Teramo, Italy - 
\email{cassisi,pietrinferni@oa-teramo.inaf.it}\\
$^2$ARI, Liverpool John Moores University, Twelve Quays House, Egerton Wharf, Birkenhead CH41 1LD, UK
\email{ms@astro.livjm.ac.uk}
}

\authorrunning{Cassisi et al.}

\titlerunning{Stellar models for multiple stellar populations}

\abstract{We briefly summarize the impact of the chemical peculiarities associated to the multiple
population phenomenon in Galactic Globular Clusters, on the evolutionary properties and spectral energy distribution
of second generation stars, in comparison with the primordial stellar component. 
}
\maketitle{}

\section{Introduction}

Since about 25 years, it is known that Galactic GCs host a fraction of stars characterized by chemical patterns which appear to be peculiar when compared with that of the bulk of the cluster stellar population. However, in this last decade the possibility of performing accurate multi-objects spectroscopy has clearly shown that what seemed to be an anomaly, was indeed a normal behaviour among GC stars. More in detail, extensive spectroscopical surveys of several GCs have shown the existence of well-defined chemical patterns among stars within individual clusters, as the existence of light-elements (anti-) correlations - the most famous being Na-O anti-correlation (Gratton, Carretta \& Bragaglia~2012). These empirical findings have challenged the idea that GCs are the prototype of Simple Stellar Populations (SSPs), i.e. they host coeval stars formed with the same initial chemical composition. 
This notwithstanding, the damning evidence supporting the presence of multiple stellar populations in a GC has been provided by high-precision HST photometry.  In fact, accurate photometric investigations have revealed the existence of multiple 
Main Sequence (MS) and/or Sub-Giant Branch (SGB) and/or Red Giant Branch (RGB) sequences in the CMD of various GCs. Actually the features observed in the CMD change significantly from one cluster to another, and their properties do strongly depend on the adopted photometric systems (see Milone et al.~(2010), and G. Piotto in this volume). 

The commonly accepted scenario is that, in any GC, a second (and in some cases also more) generation(s) of stars can form from the ejecta of intermediate-mass and/or massive stars belonging to the first stellar population formed 
during the early phase of the cluster evolution, whose chemical composition is modified by high-temperature proton captures. 
Although some amount of dilution between pristine (unpolluted) matter and nuclearly processed matter seems to be unavoidable in order to explain many observational evidence, the second generation stars would be formed via matter characterized by light-element (anti-)correlations and He enhancement. 

The ability to trace both spectroscopically and photometrically the various sub-populations hosted by each individual GC, allows 
now the identification of both the primordial stellar component (the first generation, FG) and the second generation (SG) stars.
The existence of a strong correlation between the spectroscopic signatures of the distinct sub-populations and their 
distribution along the multiple CMD sequences, has suggested that the peculiar chemical patterns of SG stars have to affect both the
evolutionary properties of these stars as well as their spectral energy distribution.

To date, investigations have provided clear evidence that the multiple CMD sequences can be interpreted as due to  \lq{quantized}\rq\ He abundances, as in the case of $\omega$~Cen and NGC~2808, distinct CNO abundance pattens as in the case of M~22 and (the still-debated case of) NGC1851, and the presence of light-element (anti-)correlations. 
To trace correctly the various sub-populations from CMD analyses, stellar models must properly account 
for the observed chemical patterns in both FG and SG stars.

In the last decade, a substantial effort has been devoted to investigate the effect 
of the chemical patterns characteristic of the multiple population 
phenomenon on both stellar models and model atmospheres, and the corresponding evolutionary tracks, isochrones and 
colour - ${\rm T_{eff}}$ transformations. 
In this contribution we discuss how these chemical peculiarities affect the evolution of SG stars, as well as 
their photometric properties. 

\section{Theoretical framework}

\subsection{He-enhancement}

Helium is one of the most important chemical species in the context of stellar evolution, because any change of its abundance hugely affects the structural and evolutionary properties of stars. In more detail, a change of the He abundance affects low-temperature, radiative opacities, for an increase of He causes a reduction of the opacity. This is due to the fact that when increasing the helium abundance at fixed metallicity, the H abundance has to decrease, and given that H is a major opacity source via the ${\rm ^-H}$ ion, this causes a global reduction of the opacity. This effect explains why He-enhanced stellar models have hotter ${\rm T_{eff}}$ values in comparison with normal He abundance stars. The opacity reduction due to He-enhancement contributes also to make brighter the stars during the core H-burning stage, although the larger contribution to the change in the stellar surface luminosity is due to the change in the mean molecular weight associated to the He abundance increase. In fact, the H-burning efficiency is strongly dependent on the value of the mean molecular weight $\mu$: ${\rm L_H\propto\mu^7}$. When He increases the mean molecular weight (at fixed metallicity) has to increase and this translates in a larger H-burning efficiency. Given that 
${\rm \Delta{L_H}/L_H=7\Delta\mu/\mu}$, a change $\Delta{Y}=0.10$ --  lower than the maximum He enhancements expected for the bluest MS stars in GCs $\omega$~Cen and NGC~2808 -- causes a $\sim 50$\% variation of the H-burning efficiency compared to normal-He stars. The combined effect of radiative opacity decrease and H-burning efficiency increase causes He-rich stellar models to be brighter and hotter during the MS stage. As a consequence, their core H-burning lifetime ($t_H$) is significantly reduced: for a $0.8M_\odot$, $t_H$ decreases by $\sim48$\% when increasing the He content from the primordial value Y=0.245 to 0.40. 

\begin{figure}[]
\resizebox{\hsize}{!}{\includegraphics[clip=true]{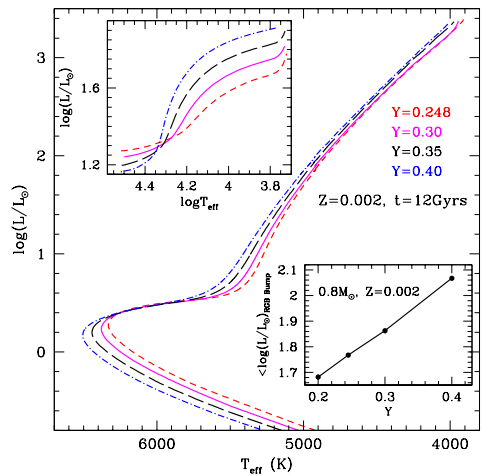}}
\caption{\footnotesize
Comparison between 12~Gyr-old, Z=0.002, isochrones computed for various assumptions about the initial He abundance. The two insets
shows the location of the ZAHB, and the trend of the average RGB bump brightness as a 
function of the initial He abundance for a ${\rm 0.8M_\odot}$.}
\label{fig_elio}
\end{figure}

Fig.~\ref{fig_elio} shows different isochrones computed with the same [Fe/H] and age, but various values for the initial He content. There are some interesting features disclosed by this plot: {\sl i)} the He-rich MS runs parallel in the luminosity interval from the MS Turn-off (TO) and the location of stars with mass $\sim0.5M_\odot$; {\sl ii)} the MS effective temperature is sensitive to the He increase (${\Delta{T_{eff}}}/{\Delta{Y}}\approx 2.3\times10^3 K$); {\sl iii)} at fixed age, the mass at the MS TO significantly decreases when increasing the initial He content. For a 12Gyr old isochrone it is equal to $0.806M_\odot$ for Y=0.245 and $0.610M_\odot$ for Y=0.40;   {\sl iv)} for any given age, the SGB is not affected by a He change; {\sl v)} the ${\rm T_{eff}}$ values of RGB stellar models are also affected by an He increase, although to a smaller extent than the MS locus. 
The predicted behaviour of the MS as a function of the initial He abundance justifies the interpretation of the MS splitting observed in some GCs, like $\omega$~Cen and NGC~2808.

When considering the RGB evolution, there are two important features that are affected by a change of the initial He abundance: the RGB  bump and the brightness of the RGB tip (TRGB). Model computations predict: {\sl a)} an increase of the bump brightness (see Fig.~\ref{fig_elio}), and {\sl b)} a smaller luminosity excursion during the RGB bump stage. The first effect is due to the lower envelope opacity of He-rich stars, that causes the discontinuity of the H abundance left over by the {\sl first dredge-up} to be located in more external layers. As a consequence, the H-burning shell encounters the discontinuity at later times, hence at a brighter luminosity. The second effect is caused by the fact that in He-rich stars, the jump of the H abundance at the discontinuity is smaller than in normal He stars. As a consequence the surface stellar luminosity is less affected when the H-burning shell crosses the discontinuity. From an observational point of view, the impact of an He enhancement on the RGB bump brightness in GCs has been discussed by Bragaglia et al.~(2010) . On the other hand, the effect on the RGB bump luminosity excursion, hence evolutionary lifetime, has been used by Nataf et al.~(2011) to interpret the anomalously 
small number of RGB bump stars in the Galactic bulge. A number smaller than predictions by He-normal stellar models has been considered a proof that bulge stellar populations are He-enhanced.

The TRGB brightness is significantly affected by an He-enhancement. For a given total stellar mass increase of the initial He content 
decreases the TRGB brightness. This is due to the fact that the interiors of He-rich stars are hotter at the end of the core H-burning stage, and when they reach the RGB stage they develop a significantly lower level of electron degeneracy in their He core. At the same time, as a consequence of the larger H-burning efficiency, the He core mass grows at a  faster rate. Both effects make \lq{easier}\rq\ to achieve the thermal conditions required by the $3\alpha$ reaction ignition, and He ignition is attained with 
a smaller He core mass. Due to the existence of a {\sl He core mass - luminosity} relation for RGB stars, the TRGB brightness decreases in He-rich, low-mass giants. Unfortunately, due to the low number of stars populating the brighter portion of the RGB and the still unsettled issue of the GC distance scale, it is almost impossible to verify observationally this prediction. 
We also note that, as a consequence of the reduction of the value of $t_H$ for He-rich stars, 
the mass of stars at the TRGB is expected to be significantly smaller in He-enhanced stellar populations, when age is kept constant 
(and if RGB  mass loss does not have a strong dependence on He). 
This occurrence explains why He-enhanced stellar populations are characterized by a bluer Horizontal Branch (HB) morphology with respect to a FG stellar population.

When moving to the core He-burning stage, one notices that Zero Age HB (ZAHB) brightness is a strong function of the initial He content. When Y increases, the ZAHB becomes brighter for ${\rm T_{eff}}$ lower that $\sim20000$K, and fainter at higher  ${\rm T_{eff}}$ 
(see Fig.~\ref{fig_elio}). This behaviour is the consequence of both the decrease of the He-core mass at the TRGB, and increased efficiency of the shell H-burning in He-rich stars. In ZAHB objects cooler than $\sim20000$~K, the second effect dominates over the first one and the stars appear brighter, whereas in the hottest portion of the ZAHB -- due to the tiny envelope mass -- the H-burning shell is not efficient enough, and the decrease of the He core mass is the dominating effect. This has the important consequence that the slope of the ZAHB in the H-R diagram (as well as in the various observational planes) is strongly dependent on the (spread in the) initial He abundances of the various sub-populations hosted within a given GC. This theoretical prediction can provide a direct explanation for the existence of a tilted HB in GCs like NGC6388, NGC6441 and NGC1851. He-enhancement has an additional important implication for the global HB morphology. Due to the combined effect of the lower RGB evolving mass and more extended blue loops that characterize the off-ZAHB evolution of He-rich stars, for a given average efficiency of the mass loss along the RGB the predicted HB location of He-enhanced models will be on average hotter, i.e. bluer, than that of He-normal stars. This helps explaining the existence the very blue HB morphology -- as well as the presence of extended HB blue tail -- in GCs hosting He-enhanced sub-populations.

\begin{figure*}[t!]
\resizebox{\hsize}{!}{\includegraphics[clip=true]{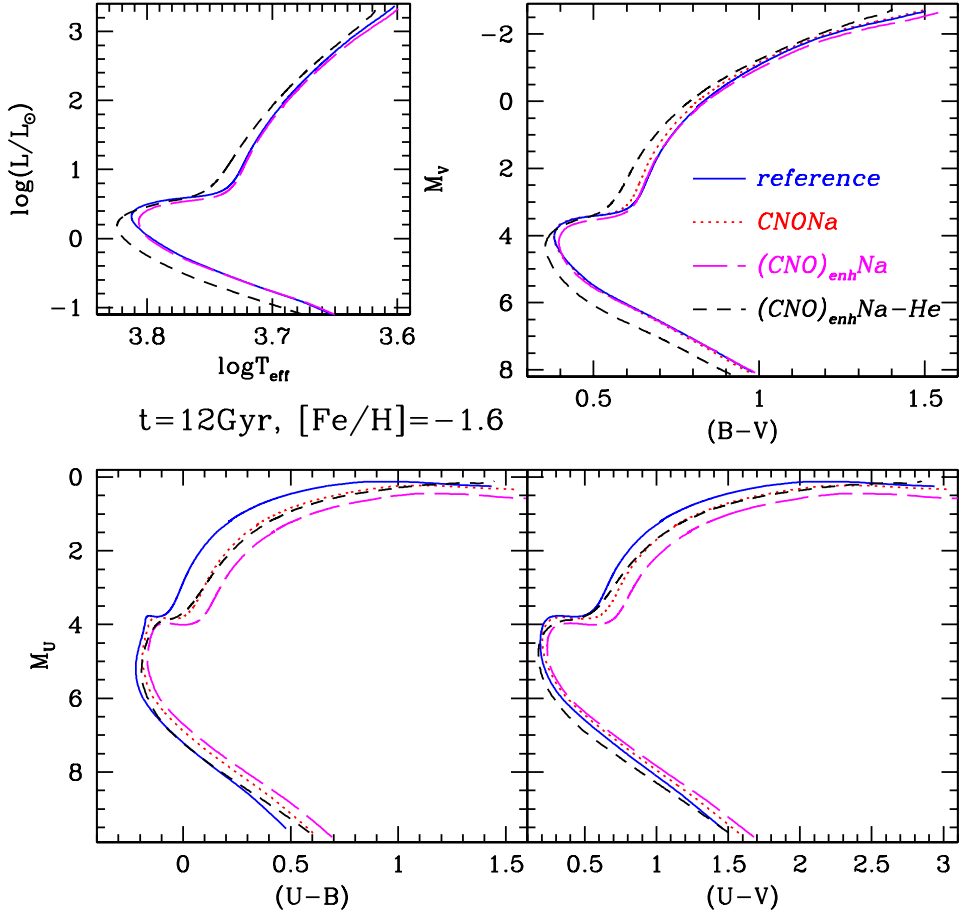}}
\vskip -1.0truecm
\caption{\footnotesize
Theoretical isochrones for the same age and metallicity (see label) but for various assumptions about the light-element distribution and
He-enhancement. More in detail, {\sl reference}: canonical $\alpha-$enhanced mixture, Y=0.248; {\sl CNONa}: light-element (anti-)correlation but the same CNO sum and He abundance of the $\alpha-$enhanced composition; $(CNO)_{enh}Na$: as before but the CNO sum is now enhanced by a factor $\sim2$; $(CNO)_{enh}Na-He$: as before but now the initial He abundance is equal to Y=0.40.
}
\label{fig_anti}
\end{figure*}

\subsection{Light-element (anti-)correlations}

Light element abundance changes can affect the stellar properties via the effects induced on both the radiative opacity evaluations and -- at least in the case of C, N and O, i.e. the CNO-cycle catalysts -- the H-burning efficiency. Many investigations have been devoted to this topic (Salaris et al.~2006, Cassisi et al.~2008, Pietrinferni et al.~2009, Ventura et al.~2009, Vandenberg et al.~2012) and the main results can be summarized as follows:

\begin{itemize}

\item even in case of extreme light-element anti-correlations, as long as the sum of CNO elements is kept constant, the evolution in the H-R diagram is unchanged compared to standard $\alpha-$enhanced stellar models;

\item if the CNO sum is enhanced, the morphology of the evolutionary tracks in the H-R diagram is modified.  
For a given iron content and stellar mass, the SGB appears fainter in comparison with standard $\alpha-$enhanced models. 
This behaviour is mainly due to the fact that the efficiency of the CNO-cycle increases when the CNO sum is enhanced. 
The ${\rm T_{eff}}$ scale for both MS and RGB CNO-enhanced stellar models is marginally affected (by less than 20~K);

\item when computing isochrones, unless the CNO sum is increased with respect the reference $\alpha-$enhanced mixture, 
the effect of the light-element (anti-)correlations is negligible.  When the C+N+O sum is increased, a separation appears along the SGB, and a CNO-enhanced
isochrone is almost perfectly mimicked by a \lq{canonical}\rq\ $\alpha-$enhanced, $\sim(1.5-2)$~Gyr older, isochrone.

\end{itemize}

Isochrones for various assumptions about the heavy element distribution and/or the initial 
He abundance, but for the same age and [Fe/H], are shown in Fig.~\ref{fig_anti};

\section{On the photometric appearance of multiple populations}

The isochrones shown in Fig.~\ref{fig_anti} reveal that in the theoretical H-R plane, the only possibility to observe a separation between isochrones with a standard $\alpha-$enhanced isochrone and with the composition of SG stars, is to assume 
a huge He enhancement (that affect MS and RGB) and/or a significant CNO-enhancement (that affect the SGB sequence). 
Needless to say that this theoretical evidence is in stark contrast with observations, that reveal the presence of splitting/broadening of the photometric  sequences (in particular, the RGB one) also in those GCs that do not show (spectroscopical) evidence of a significant CNO enhancement and/or huge He enhancement. The chemical peculiarities of second generation stars must therefore have 
an effect also on the spectral energy distribution.

Sbordone et al.~(2011) have demonstrated the crucial role played by CN and NH molecules (whose abundances in SG stars are very different from those in FG stars) in modifying the stellar spectrum at wavelengths shorter than $\sim400$~nm. This has the important implication that only magnitudes corresponding to photometric filters bluer than the standard Johnson B filter are affected by the peculiar chemical patterns of multiple populations; as a consequence, the photometric appearance of these sub-populations depends on the adopted photometric systems:

\noindent
{\underline{\it $BVI$ CMDs}}: 
a splitting (or a spread) of sequences along the MS up
  to the Turn-off (TO), and to a lesser degree of the RGB can only be achieved by varying
  the helium content. The CNONa (anti-)correlations do not influence significantly 
  stellar models and spectral energy distribution, when the C+N+O abundance  is unchanged.
On the other hand, a variation of the CNO sum with respect the \lq{canonical}\rq\ value leads to a split of the SGB; this is a 
purely evolutionary effect. 

\noindent
{\underline{\it $UBV$- and  $uy$ Str{\"o}mgren CMDs}}:
(anti-)correlations in CNONa abundances as well as  He differences may
lead to multiple sequences from the MS to the RGB, where the effect
tends to be larger, and may reach 0.2--0.3~mag. This multiplicity is
independent of the CNO sum. However, the individual element variations are decisive. 
We note that He enhancement works in the opposite direction than CNONa (anti-)correlations.

\noindent
{\underline{\it $vy$ CMDs}}: as in the case of the $BVI$-colours, a splitting of the MS up to
  the TO can be achieved only by varying the He abundance. A split of the SGB is the result of a
  change in the C+N+O abundance. Additionally, a split along the RGB may result both from both helium
 and C+N+O variations; this is different from the $BVI$-case.
 
\noindent
{\underline {\it $m_1$uy CMDs}}:  light element (anti-)correlations lead to splits along the MS, the SG and RGB; also helium variations lead to colour differences. However, the sign of the colour change is different for the lower and upper part of the RGB.

\noindent
{\underline{\it $c_y$V CMDs}}:
here, all the evolutionary sequences in the CMD show the influence of both element
anticorrelations and helium variations, and a large separation between the various sequences can be easily achieved. It is worth 
noticing that recently a new photometric index ${\rm C_{UBI}=(U-B)-(B-I)}$ has been defined (M. Monelli, this volume) as a very powerful tool for tracing various sub-populations along the RGB. This photometric index is based on broadband photometric filters and, hence, is quite more efficient than ${\rm c_y}$, based on narrowband filters.
Model predictions about the ${\rm c_{UBI}}$ index qualitatively agree very well with the observational findings.
 
\begin{acknowledgements}

SC warmly thanks the organizers for inviting him to this interesting and pleasant conference, and wishes to express his gratitude to Franca for her friendship and for her painstaking effort as {\sl "Lens Maker..."}. Financial support from PRIN INAF 2012 (PI: E. Carretta) and PRIN MIUR 2010-2011, project ``The Chemical and Dynamical Evolution of the Milky Way and Local Group Galaxies'', prot. 2010LY5N2T " is acknowledged.

\end{acknowledgements}

\bibliographystyle{aa}

\end{document}